\documentclass{aa}
\usepackage[utf8]{inputenc}
\usepackage[T1]{fontenc}
\usepackage[english]{babel}
\usepackage{natbib}
\usepackage{graphicx, float, subfigure, blindtext}
\usepackage{xcolor}
\usepackage{mathtools}
\usepackage{multirow}
\usepackage{tabularx}
\usepackage{float}

\usepackage{bm}
\graphicspath{{images/}}
\usepackage{MR_AA}

\newcommand\perio{{\tt Period04}}

\title{Seismology of Altair with MOST\thanks{This work is based on data from the MOST satellite, a Canadian Space Agency mission, jointly operated by Dynacon Inc., the University of Toronto Institute for Aerospace Studies and the University of British Columbia, with the assistance of the University of Vienna.}}

\author{Cécile Le Dizès\inst{1,2}, Michel Rieutord\inst{1} and Stéphane Charpinet\inst{1}}
\date{\today}
\institute{
IRAP, Universit\'e de Toulouse, CNRS, UPS, CNES,
14, avenue \'{E}douard Belin, F-31400 Toulouse, France
\and
Institut Supérieur de l’Aéronautique et de l’Espace (ISAE-SUPAERO), Université de Toulouse, F-31400 Toulouse, France
\\
\email{cecile.le-dizes@student.isae-supaero.fr, [Michel.Rieutord, Stephane.Charpinet]@irap.omp.eu}
}

\begin{document}

\abstract
{Altair is the fastest rotating star at less than 10 parsecs from the Sun. Its precise modelling is a landmark for our understanding of stellar evolution with fast rotation, and all observational constraints are most welcome to better determine the fundamental parameters of this star.}
{We wish to improve the seismic spectrum of Altair and confirm the $\delta$-Scuti nature of this star.}
{We used the photometric data collected by the Microvariability and Oscillations of STars (MOST)
satellite in the form of a series of Fabry images to derive Altair light curves at four epochs, namely in 2007, 2011, 2012, and 2013.}
{We first confirm the presence of $\delta$-Scuti oscillations in the
light curves of Altair.  We extend the precision of some eigenfrequencies
and add new ones to the spectrum of Altair, which now has 15 detected
eigenmodes. The rotation period, which is expected at $\sim$7h46min from
models reproducing interferometric data, seems to appear in the 2012 data
set, but it still needs confirmation. Finally, Altair modal oscillations
show noticeable amplitude variations on a timescale of 10 to 15 days,
which may be the signature of a coupling between oscillations and thermal
convection in the layer where the kappa-mechanism is operating.}{The
Altair oscillation spectrum does not contain a large number of excited
eigenmodes, which is similar to the fast rotating star HD220811. This
supports the idea that fast rotation hinders the excitation of eigenmodes
as already pointed out by theoretical investigations.}

\maketitle

\section{Introduction}

\object{Altair} ($\alpha$ Aquilae, \object{HD187642}) is one of the few
fast rotating early-type stars in the solar neighbourhood, its distance
being 5.13~pc according to HIPPARCOS data \citep{vanleeuwen07}. As
such, it is an ideal target of interferometric observations,
which have regularly measured its shape and surface brightness
\citep{vanbelle+01,domiciano+05,monnier_etal07}.  These observations have
led to the determination of the centrifugal flattening of Altair which
turns out to be close to 22\% while its angular velocity at the equator
is 74\% of the Keplerian one. Such a centrifugal distortion makes the
use of two-dimensional models mandatory for a correct interpretation
of observational data. \cite{bouchaud+20} actually performed the
first 2D modelling of Altair using ESTER 2D models which include,
self-consistently, the 2D structure and the large-scale flows, namely
differential rotation and meridional circulation, driven by baroclinicity
\citep{ELR13,RELP16}. Besides demonstrating the young age of Altair
($\sim100$~Myrs instead of 1~Gyrs as previously estimated with 1D models
e.g. \citealt{domiciano+05}), the work of \cite{bouchaud+20} shows us that
the seismic spectrum of Altair is a key ingredient to further constrain
the mass of this star. Indeed, acoustic modes are sensitive to the mean
density \citep{reese+12,garciahernandez_etal15}, while interferometry,
through the measurement of radii, determines the volume of the star.

Hence, the detection of (presumably) acoustic oscillations by \cite{buzasietal05}, in a photometric monitoring of Altair in 1999 by the star tracker of the WISE mission, was a good surprise. This detection meant that Altair is a $\delta$-Scuti star, and actually the brightest one \citep{buzasietal05}. However, these oscillations were never confirmed. Regarding their importance in the modelling of this star, we looked for new data that would confirm and improve the result of \cite{buzasietal05}. Altair has been quite intensively observed with the  Microvariability and Oscillations of STars (MOST) satellite \citep{walker+03} in 2007, 2011, 2012, and 2013, but no analysis of these data has been published so far. Since such data have the potential to confirm the previous results and possibly show new frequencies or variations of modes amplitudes, we embarked on a project to process them in order to once again exhibit the seismic spectrum of this fascinating star.

This paper is organised as follows: We first describe the data and the processing we applied to extract the light curves (sect. 2). We then analyse the light curves and retrieve the oscillation spectra at the various epoch of the data (sect. 3). A discussion and preliminary conclusions end the paper.

\section{Observations and data reduction}

Altair was observed in four sequences by the MOST satellite, which
run around the Earth at a mean altitude of 830~km with an orbital
period of 6084.9~s, corresponding to a frequency of 14.199 c/d
\citep{walker+03}. The characteristics of the time series are summarised
in Tab.~\ref{tab:the_data}.

Data were downloaded from the Canadian Astronomy Data Centre. They
are available as a series of Fabry images, from which we derived light
curves. In Appendix A, we provide some details on the first steps of
this processing, but the main challenge of this data reduction is the
clean suppression of stray light, which comes from the illuminated side
of the Earth or from moonshine.

\begin{figure}
    \centering
    \includegraphics[width=\linewidth]{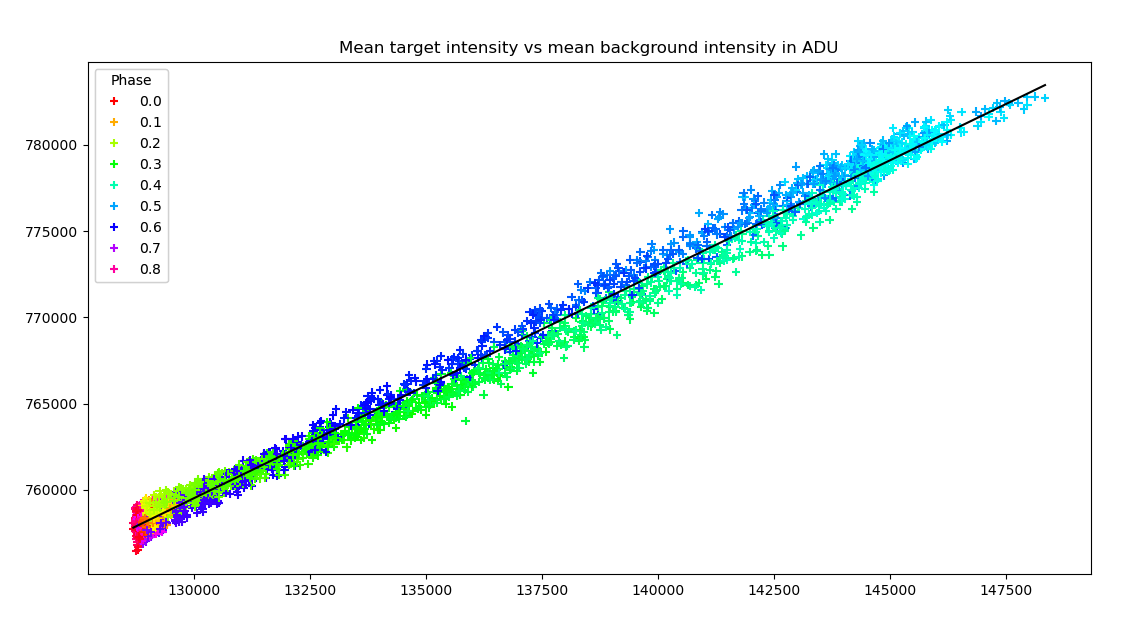}
    \caption{Correlation between target and background pixels. Colours indicate the orbital phase of MOST.}
    \label{fig:correlation}
\end{figure}

\begin{figure}[t]
    \centering
    \includegraphics[width=\linewidth]{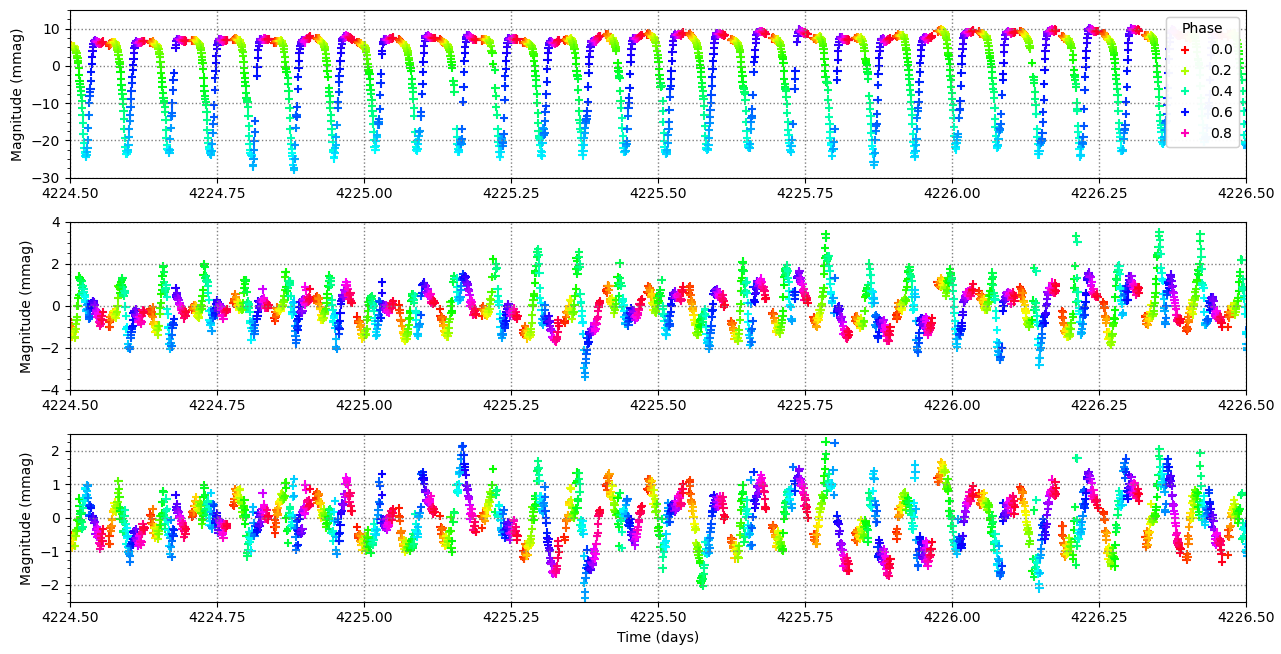}
    \caption{
Effects of the different reduction steps on the magnitude amplitude for the 2011 set. Top: Original data.\ Middle: Data after the decorrelation step. Bottom: Data after the decorrelation and removal of the mean orbit. Colours indicate the orbital phase of MOST.}
    \label{fig:step_reduction}
\end{figure}

The technique we used is inspired by the one developed by \cite{huber_reegen08} and is based on calculating correlations of mean target and background flux. Since stray light is variable over timescales of approximately a day, the correlation was calculated using the mean background and target intensity in a moving window of 1 day rather than using the whole light curve. 
An example of a linear correlation between target and background pixels is shown for the 2011 data set in Fig. \ref{fig:correlation}.

This method assumes that the influence of stray light is the same for target and background pixels, and that the background pixel intensity is not influenced by the star light.

The correction of the correlation was made by measuring the variations $\delta m$ between the real target magnitude and the one given by the linear trend each time. Hence,

\begin{equation}
    \delta m (t)= -\frac{5}{2\ln{10}}\ln\lc\frac{I_{tar, real}(t)}{I_{tar, correl}(t)}\rc
,\end{equation}
where $I_{tar,real}(t)$ is the mean intensity of the target pixels at time $t$, and

\[I_{tar,correl}(t) = aI_{bck,real}+b,\]
where $I_{bck,real}(t)$ is the mean intensity of the background pixels at time $t$, and $a$, $b$ are the coefficients given by the linear regression.

\begin{figure}[t]
    \centering
    \includegraphics[width=0.95\linewidth]{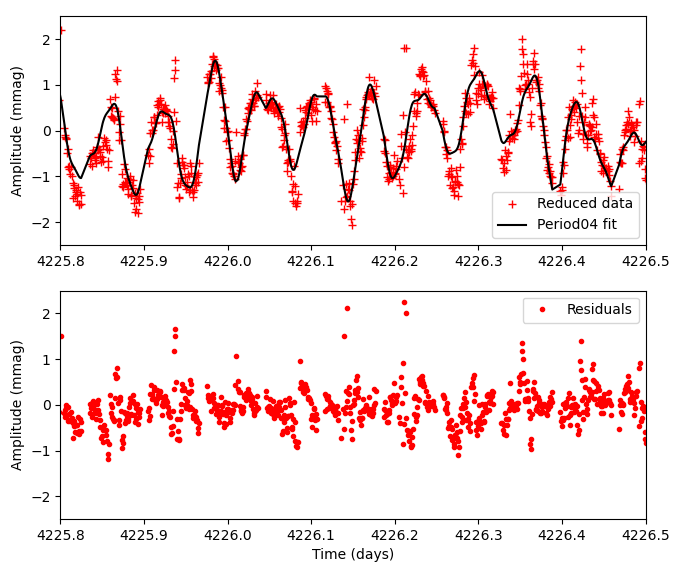}
    \caption{Amplitude (in mmag) versus time (in days) for a portion of the 2011 light curve. Top: Reduced data with the solid line showing the fit given by Period04.
    Bottom: Residuals after subtraction of the fit.}
    \label{fig:magnitude_fit}
\end{figure}
As discussed in \cite{huber_reegen08}, our procedure is a simplified
version of the algorithm of \cite{reegen+06}, which searches for the
background pixels that best correlate with each target pixel. In the
procedure of \cite{reegen+06}, the
background flux is removed from the target pixel intensity and
the procedure is repeated on the next target pixel until all stray light
contributions have been reduced to an acceptable level.


Such a procedure is able to remove a non-uniform stray light pattern,
but it is quite demanding in computation time. We did not implement this
type of processing since, in the end, the simple algorithm described
previously, with some additional processing detailed below, satisfactorily
removed the periodic signal coming from stray light.

Hence, to further remove the effects of the orbital period at a frequency of 14.199 c/d, we used a moving window of 28 periods ($\sim 2~d$) to create a mean light curve folded at the MOST orbital period and removed this mean light curve from the data. This step effectively suppressed the harmonics of the orbital frequency from the final power spectra.

\begin{table*}[t]
    \centering
    \begin{tabular}{cc|c|c|c||c|c|c|}
    \multicolumn{2}{c}{}&\multicolumn{3}{|c||}{2011} & \multicolumn{3}{c|}{2012} \\
    \cline{3-8}
    &&&&&&& \\
    \multicolumn{2}{c|}{}&\perio & FELIX & PYPE & \perio & FELIX & PYPE \\
    &&&&&&& \\
    \hline 
    &&&&&&& \\
    \multirow{2}{*}{Mode 1}&F&15.7677 $\pm$ 46& 15.7657$\pm$37 & 15.7812$\pm$16 &15.7685$\pm$3 & 15.7686$\pm$4 & 15.7689$\pm$1\\
    &A&585$\pm$12 & 564$\pm$15 & 662$\pm$11 &590$\pm$19 & 583$\pm$15 & 583$\pm$7\\
    &&&&&&& \\
    \hline 
    &&&&&&& \\
     \multirow{2}{*}{Mode 2}&F&20.7898$\pm$ 60& 20.7832$\pm$81 & 20.7898$\pm$35 &20.7865$\pm$6 & 20.7866$\pm$9 & 20.7863$\pm$4\\
    &A&241$\pm$10 & 246$\pm$14 & 245$\pm$7 &262$\pm$8 & 262$\pm$14 & 262$\pm$6\\
    &&&&&&& \\
    \hline 
    \end{tabular}
    \bigskip
    \caption{Comparison of various software (\perio\ , FELIX, PYPE) at determining the frequency and amplitude of two modes with different difficulty. The 2011 data are short (4 days long) and thus less resolved than the 2012 data (33 days long). The 15.77 c/d is close to two other frequencies, while the 20.79 c/d mode is isolated. Frequencies (F) are in c/d, their uncertainties are in 10$^{-4}$c/d, while amplitudes (A) are in ppm as their uncertainties.}
    \label{compar}
\end{table*}

\begin{figure*}[t]
    \centering
    \includegraphics[width=0.99\linewidth]{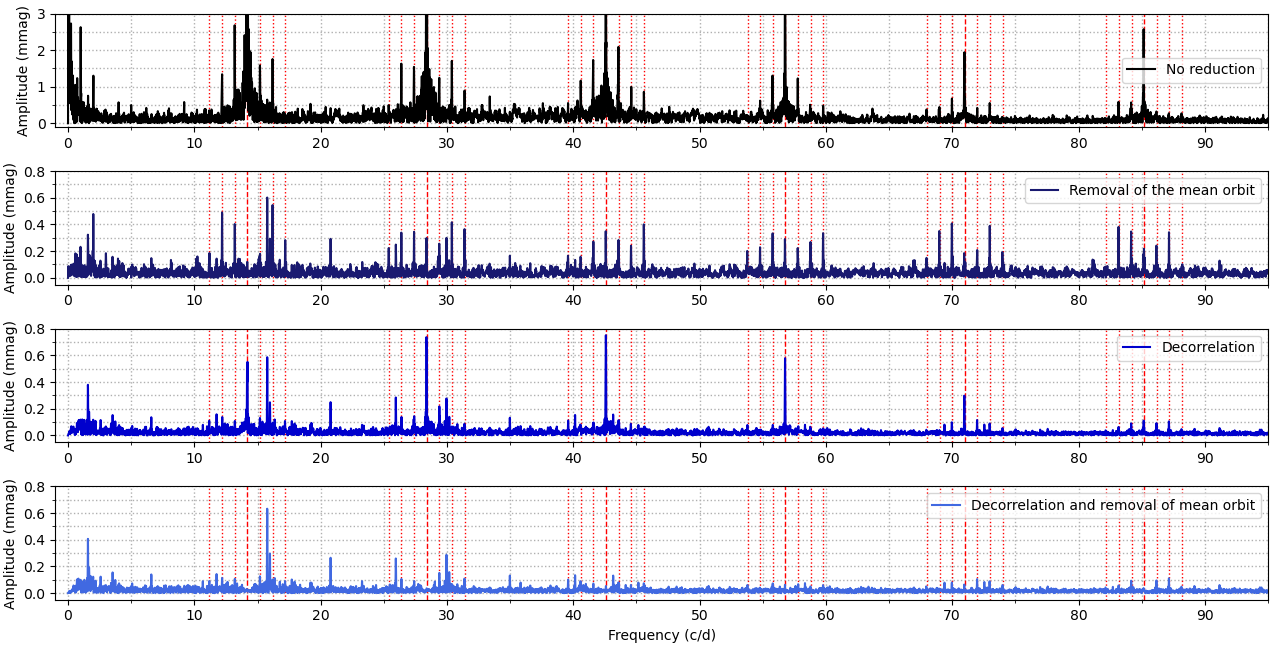}
    \caption{Effects of different reduction techniques on the amplitude spectrum for the 2012 set. The dashed red lines show the orbital frequency and its harmonics. The red dotted lines are evenly spaced 1c/d apart from one another.}
    \label{fig:processing_sp2012}
\end{figure*}

To illustrate this processing, we show in Fig.~\ref{fig:step_reduction} the evolution of the light curve at each step using a subset of the 2011 data. The final light curve, a glimpse of which is shown in Fig.~\ref{fig:magnitude_fit}, typically gives amplitude fluctuations of $\pm2$~mmag. This is similar to those observed by \cite{buzasietal05} and also quite similar to those of Rasalhague ($\alpha$ Ophiuchi), another fast rotating A-type star with $\delta$-Scuti oscillations \citep{monnier_etal10}.

Fig.~\ref{fig:processing_sp2012} further shows the role of the
processing steps on the power spectra of the light curves using Period04
\citep{lenz_breger05}. The orbital frequency at 14.199~c/d clearly shows
up in the unprocessed data, as well as in the decorrelated data (at a
lower level of course). We also notice secondary peaks separated from
the orbital frequency by $\delta\nu=1$~c/d or multiples of it, namely
associated with Earth's rotation.  Fig.~\ref{fig:processing_sp2012}
clearly shows that the removal of the mean orbit signal and global
decorrelation quite nicely suppresses most of the systematics coming
from the orbital motion of MOST.

\section{Analysis}
\subsection{Algorithm - Validation}

After the foregoing extraction of the light curves from the raw data,
we proceeded with the derivation of the frequencies and amplitude of the
modes. For that purpose, we used several types of software that are all
based on the same classical method, involving computing the Lomb-Scargle
periodogram, pre-whitening the signal, and adjusting a combination of
sinusoidal signal as

\begin{equation}
I(t)=\sum_{n=1}^{N}A_n\sin(2\pi\nu_nt+\phi_n)
\label{the_model}
\end{equation}
to the data through non-linear least-square optimisation. We used \perio\
\citep{lenz_breger05}, FELIX \citep{charpinet2010,zong2016}, and PYPE,
which is a python software specially built for this work from
python libraries (Astropy 4.2 for the Lomb-Scargle periodogram and
Scipy for the optimisation). These types of software differ from one
another basically from their evaluation of uncertainties. Period04 uses
Monte-Carlo simulations, FELIX uses formulae from \cite{montgomery+99},
while PYPE is based on the bootstrap method. Comparison of the results
from these three types of software is useful to appreciate the influence
of the numerical procedure on the results. Basically, the programmes
usually agree, especially when the quality of the data increases, not
surprisingly. We show in Tab.~\ref{compar} the results for two modes
using the three programmes. This illustrates that amplitudes of the main
mode are badly determined with the 2011 data due to their short length
(4 days), mostly because of the mixture of this mode with its neighbouring
one at 15.98 c/d.

\subsection{Error bars, noise}

\begin{figure}[t]
    \centering
    \includegraphics[width=\linewidth]{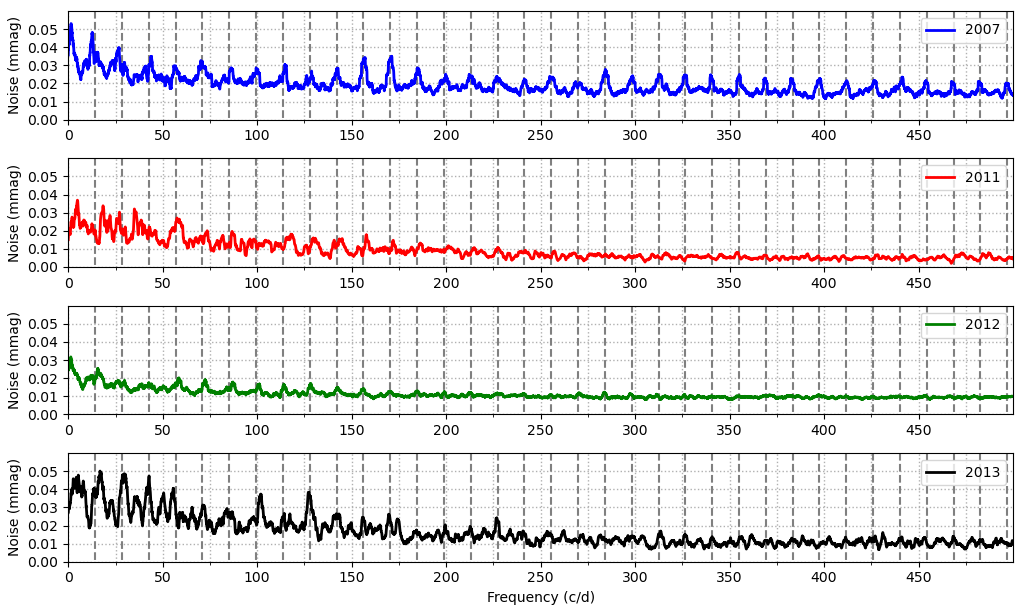}
    \caption{Noise spectrum calculated for each data set with the residuals of pre-whitened light curves with 16 frequencies.}
    \label{fig:noise_spectrum}
\end{figure}

Before presenting the results, we first discuss their sources of uncertainties. As written above, we assume that the flux variations of Altair are due to excited eigenmodes in the small amplitude regime and they can be represented by the series (\ref{the_model}). The amplitudes $A_n$, the frequencies $\nu_n$, or the phase $\phi_n$ are the parameters of the model whose uncertainties are to be evaluated. These uncertainties basically have three sources: (i) residual stray light (ii), random noise (iii), and time sampling of the signal.

As far as residual stray light is concerned, it is characterized by its frequencies that are typically of the form $F_{nm} = nf_{\rm orb}\pm mf_{\rm day}$, which we detected for $0\leq n\leq7$ and $0\leq m\leq4$. These spectral lines prevented us from detecting or correctly measuring any signal in frequency bands such as $[F_{nm}-\Delta f,F_{nm}+\Delta f]$, where $\Delta f$ is the frequency resolution of the window data set. As a consequence, Altair's frequency oscillation at 28.40 c/d$\simeq2f_{\rm orb}$ detected by \cite{buzasietal05} was just screened by the stray light signal (and the time window which includes this harmonic). The same pollution affects the oscillation at 16.18 c/d close to the frequency $f_{\rm orb}+2f_{\rm day}$.

Next, the influence of intrinsic noise was evaluated by either the
bootstrap method (PYPE), \cite{montgomery+99} analysis (FELIX), or
Monte-Carlo simulations (\perio). We note that sometimes Monte-Carlo
simulations of \perio\ do not give realistic results, especially on noisy
data. By pre-whitening the signal until the standard deviation of the
remaining light curve was approximately constant, we got an idea of the
random noise. The amplitudes of this noise for the
four data set at hand are:

\bigskip
\begin{tabular}{cccc}
    2007: & 1456\, {\rm ppm}, &2011:& 224\, {\rm ppm},  \\
    2012: & 414\, {\rm ppm}, &2013:& 374\, {\rm ppm}
\end{tabular}

\bigskip
The 2007 data set is obviously the noisiest due to large pointing errors,
but its length ($\sim$20 days) somehow compensates for this. The best data
set is the one from 2012 because of its length and moderate noise. This
is also illustrated in Fig.~\ref{fig:noise_spectrum} where we plotted
the noise spectrum for the four data sets. Clearly 2007 appears very
much polluted by stray light, while 2012 appears to be the best series.
We note that 2011 and 2013 have most of their noise in the $[0,50]$c/d
range, which is also the range of interest, unfortunately.

\begin{figure*}[t]
    \centerline{\includegraphics[width=\linewidth]{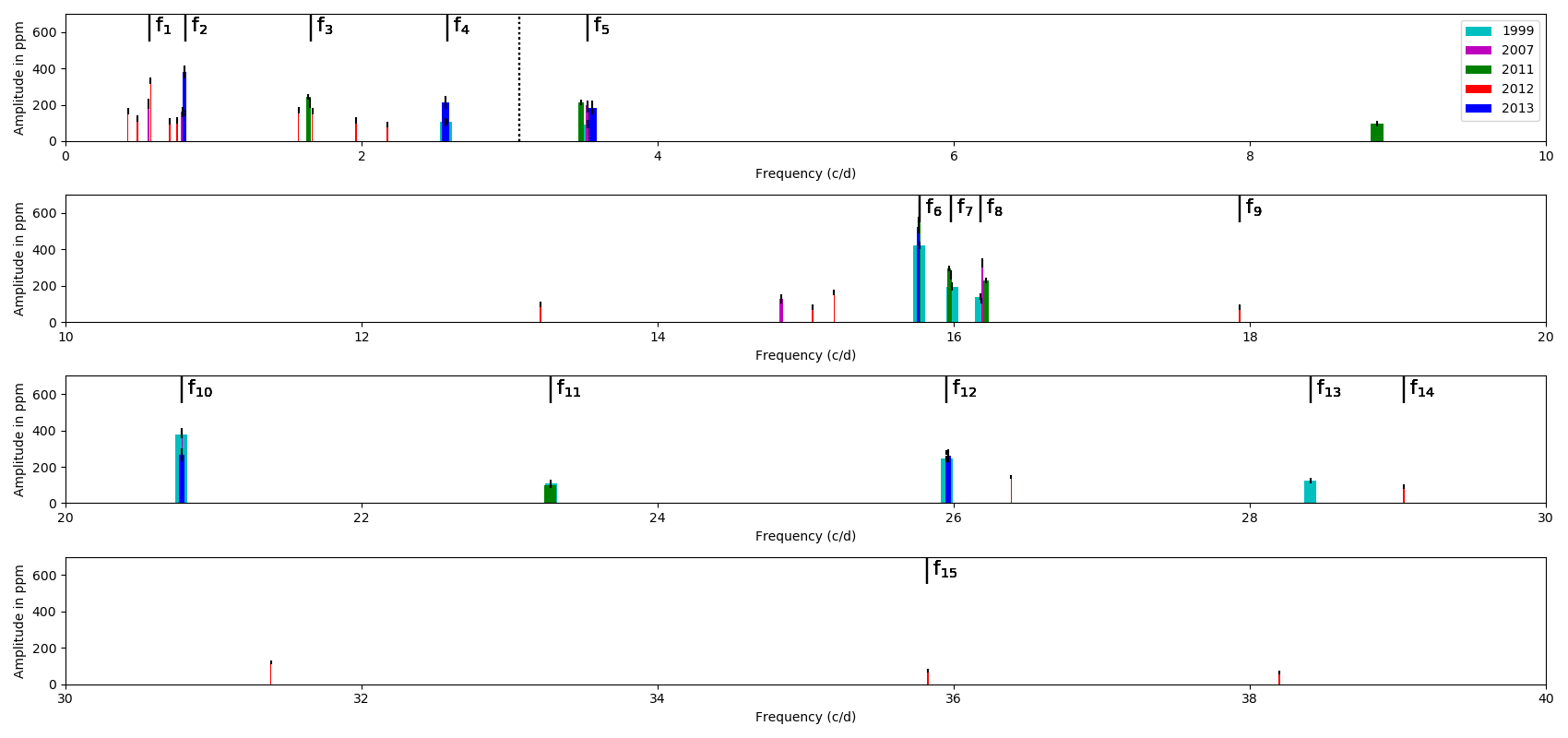}}
    \caption{Idealised amplitude spectrum of Altair derived from the analysis of the four data sets of MOST. The width of the bars shows the frequency uncertainty, while the little vertical bars over the thick previous bars show the amplitude uncertainty. Frequency labels refer to table~\ref{freq_altair}. Some unlabelled frequencies have been left and just await confirmation from future observations. The dotted vertical line indicates the expected frequency of Altair's rotation.}
    \label{fig:amplitude_spectrum}
\end{figure*}

\begin{table}[t]
    \centering
\begin{tabular}{|ll|c|l|}
 \multicolumn{2}{|c|}{Frequency} & $\moy{A}$ & Years of detection\\
\multicolumn{2}{|c|}{(c/d)}   &    (ppm)   &   \\
  & &                   &  \\
$f_1$ &  $0.57^*$     & 270   & 2007, 2012 \\
$f_2$ &  $ 0.81^*$     & 160   & 2007, 2012 \\
$f_3$ &  $ 1.66^*$     & 200   & 2007, 2012 \\
$f_4$ &2.58     & 100 & 1999, 2012, 2013 \\
$f_5$ &3.527     & 150  &  1999, 2007, 2012, 2013 \\
$f_6$ &15.7679     & 520 & 1999 $\tv$ 2013 \\
$f_7$ &15.983     & 260 & 1999, 2007, 2012  \\
$f_8$ &16.180     & 140 &  1999\\
$f_9$ &17.93$^*$  & 80 &  2012\\
$f_{10}$ &20.7865     & 330 & 1999 $\tv$ 2013  \\
$f_{11}$ &23.28     & 110 &  1999, 2007\\
$f_{12}$ &25.952     & 220 & 1999 $\tv$ 2013 \\
$f_{13}$ &28.408   &  120  &  1999\\
$f_{14}$ &29.04$^*$ & 100  & 2007, 2012\\
$f_{15}$ &35.82$^*$ & 80   & 2007, 2012\\
   \multicolumn{4}{c}{}  
\end{tabular}
\caption{List of oscillation frequencies that have been
detected at a S/N above 4. The six
starred frequencies are new and complete those detected by
\cite{buzasietal05}. Uncertainties on frequencies are typically 0.02 c/d
or better. As amplitudes vary in time (see sect.~\ref{tv}), an average
value is given.}

    \label{freq_altair}
\end{table}

\begin{figure*}[ht]
    \centering
    \includegraphics[width=\linewidth]{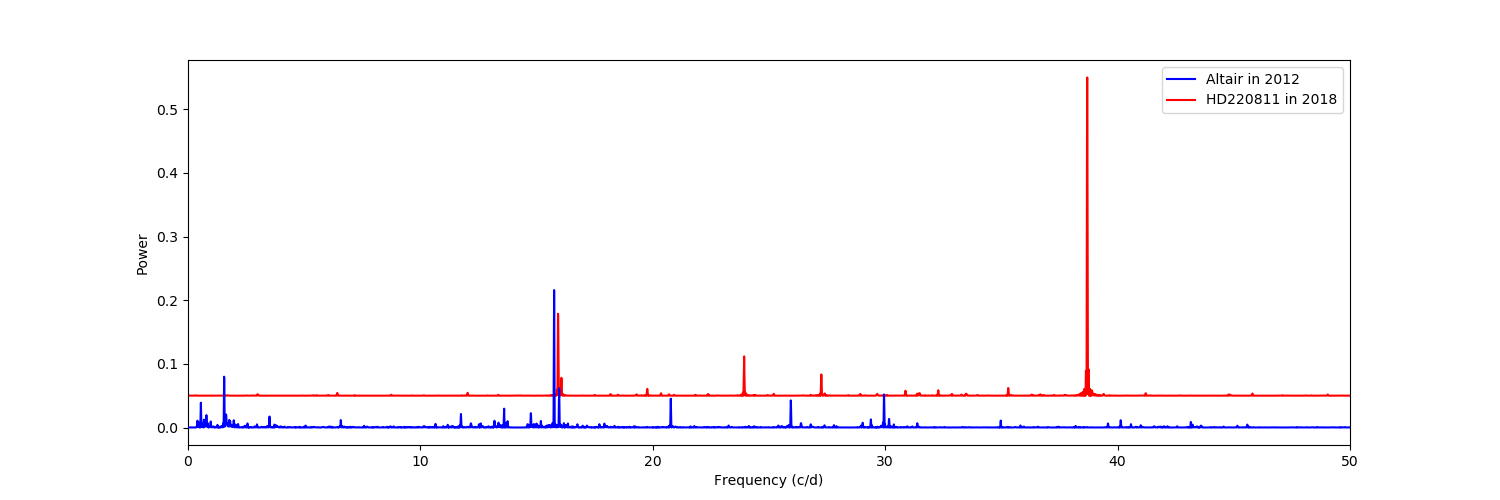}
    \caption{Superposition of Altair's and HD220811 spectra.}
    \label{HD}
\end{figure*}

\begin{figure}[t]
    \centering
    \includegraphics[width=\linewidth]{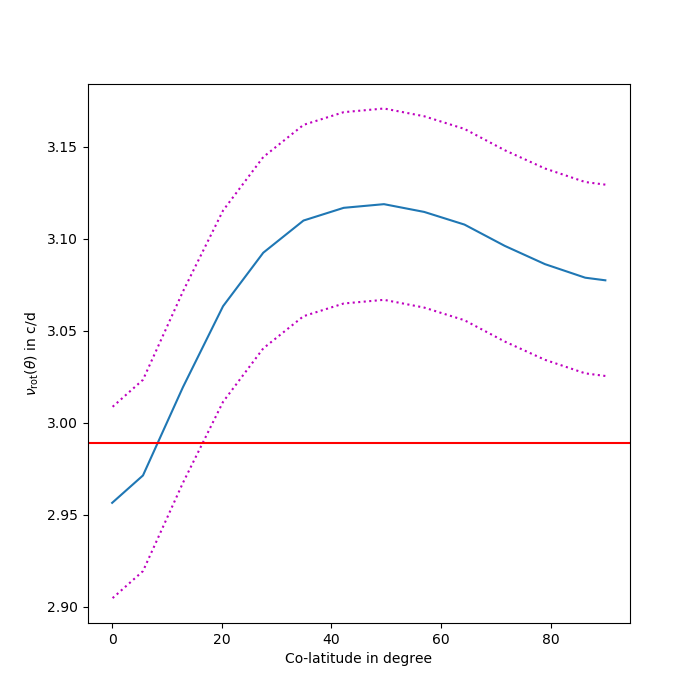}
    \caption{Surface differential of Altair from the 2D model of \cite{bouchaud+20} expressed by a frequency in c/d as a function of co-latitude (solid blue line). The dotted lines show the 2-sigma uncertainty of the model parameter, while the horizontal red line shows the frequency 2.99 c/d, which shows up in the 2012 data in Fig.~\ref{fig:Rotation}.}
    \label{fig:Diff_rot}
\end{figure}

\begin{figure}[t]
    \centering
    \includegraphics[width=\linewidth]{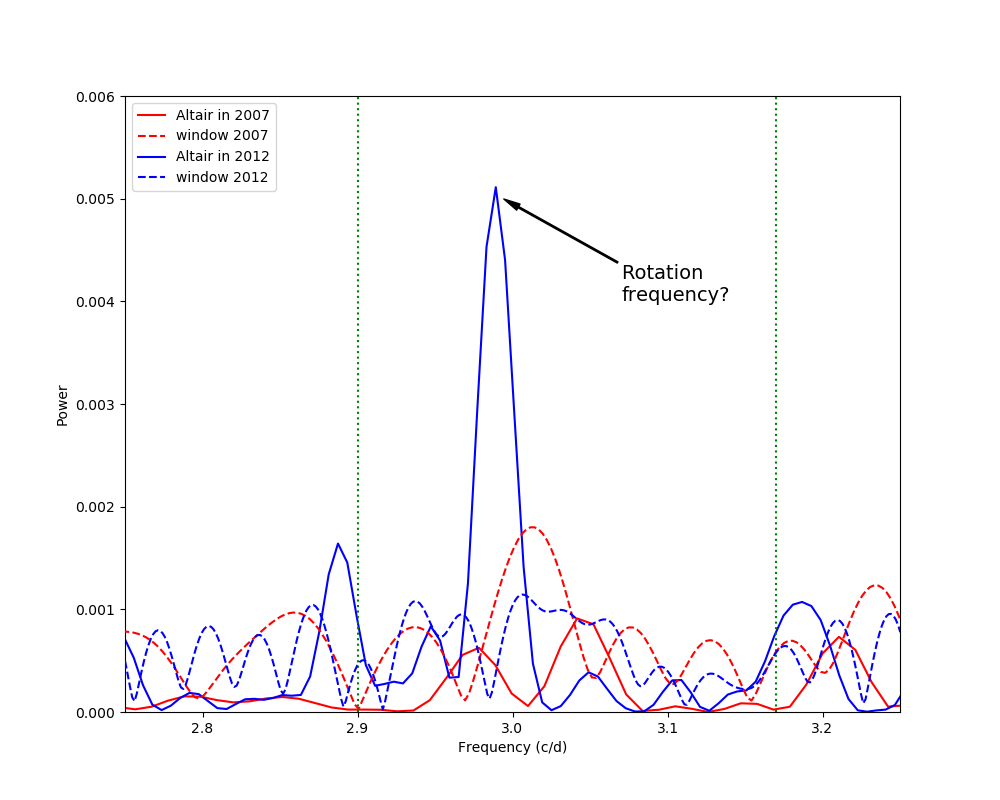}
    \caption{Close-up of the spectrum near the expected rotation frequency. The two dotted vertical lines show the 2-sigma interval where the rotation could be found including the effects of surface differential rotation.}
    \label{fig:Rotation}
\end{figure}

\begin{figure}[t]
    \centering
    \includegraphics[width=\linewidth]{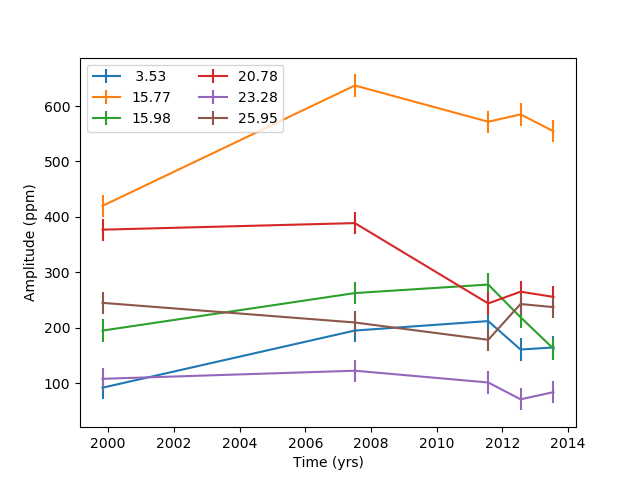}
    \caption{Time variability of the amplitudes (in ppm) of the six dominant modes with uncertainties.}
    \label{fig:Amp_vs_time}
\end{figure}

\begin{figure}[t]
    \centering
    \includegraphics[width=\linewidth]{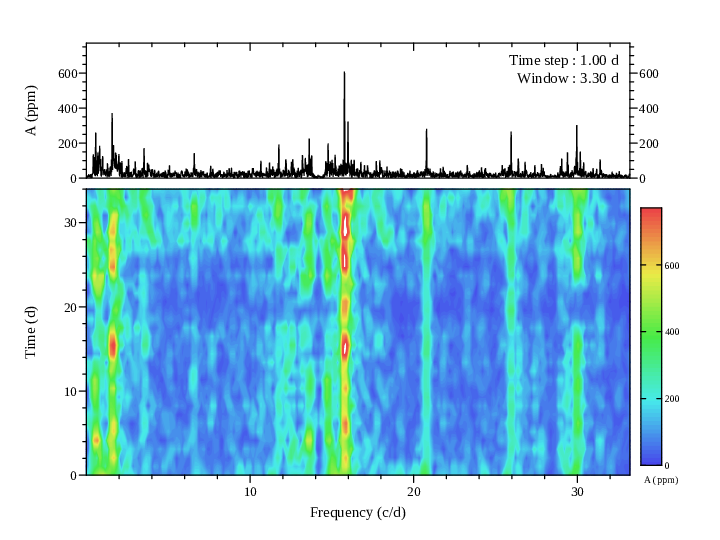}
    \caption{Time-frequency diagram for the 2012 observation set. Time starts at JD-JD2000=4576.093 .}
    \label{fig:tfd}
\end{figure}

\begin{figure}[t]
    \centering
    \includegraphics[width=\linewidth]{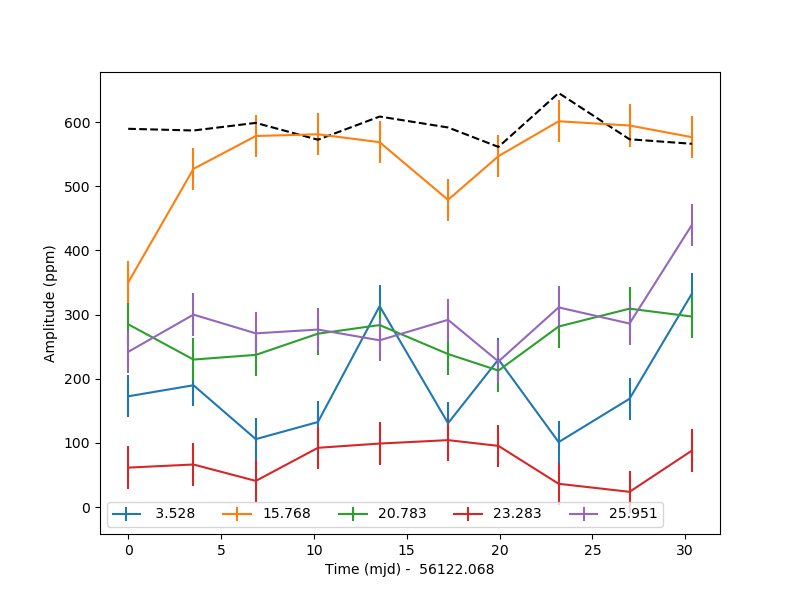}
    \caption{Time variability of the amplitudes (in ppm) of the five dominant modes with uncertainties during the 2012 observations. The black dashed line shows the amplitude variations of the $f_6=15.768$-mode if they were only due to the beats with the $f_7=15.983$-mode.}
    \label{fig:Amp12_vs_time}
\end{figure}

\section{Results}
\subsection{Altair's oscillation spectrum}

Fig.~\ref{fig:processing_sp2012} has already shown a view of the
Altair oscillation spectrum with 2012 data. However, as discussed
below, the
modes amplitude varies with time. We therefore extracted,
from each data set, the most significant frequencies along with their
amplitude and they are represented in a periodogram-like plot in
Fig.~\ref{fig:amplitude_spectrum}.  From this spectrum, we extracted a
list of modes, whose detection we consider as the most reliable. Their
frequencies and amplitudes are gathered in Tab. \ref{freq_altair}. Most
of them appear in several data sets, making their detection quite robust.

The first impression one gets looking at Altair's spectrum is
that it is rather sparse if we compare it to the well-studied
\citep{chen+16,balona14} $\delta$-Scuti \object{HD50844} observed by CoRoT
\citep{poretti+09}. Of course, HD50844 is near the TAMS and Altair is
close to the ZAMS \citep{bouchaud+20}. In addition, Altair's rotation
period is eight times shorter than that of HD50844. Altair's
oscillation spectrum better compares with that of stars picked out by
\cite{bedding+20}. These stars are young with intermediate mass showing
$\delta$-Scuti oscillations just as Altair. But unlike them, Altair
does not show modes with frequencies above 40 c/d, nor a clear regular
spacing of frequencies. Actually we might note an approximately recurrent
spacing of $\sim$2.5d$^{-1}$ with frequencies $f_6, f_{10},f_{11}, f_{12},
f_{13}$, and $f_{15}$, as already pointed out in \cite{bouchaud+20}.

Another regular spacing, $\sim0.907$c/d, may also be noticed within the low frequencies $f_2, f_3, f_4, f_5$. \cite{monnier_etal10} observed a similar pattern for two sets of modes in the oscillation spectrum of $\alpha$ Ophiuchi, with a spacing of $\sim1.71$c/d and $\sim1.74$c/d, respectively. These authors interpreted these sets of modes as the possible signature of equatorial Kelvin waves. This is an appealing interpretation, but in our case we shall wait for the spectroscopic data that are expected shortly before deciding on the meaning of these quasi-regular patterns at low frequency (Rieutord et al. 2021 in prep.).

We also observed something similar to a triplet of frequencies $f_6, f_7, f_8$ around $\nu=16$c/d with the most powerful peak of Altair's spectrum at 15.7679~c/d. The frequency spacing of 0.2~c/d of course does not correspond to a rotational splitting which would be much larger. A search in the theoretical predictions based on the 2D model of Altair computed in \cite{bouchaud+20} does not show conspicuous evidence of modes that could be identified with the detected frequencies. However, the theoretical modelling is not ripe yet, lacking the selection criterion of excited modes.

\subsection{HD220811: Potential sister of Altair}

Interestingly, Altair's spectrum
shows some similarities with that of \object{HD220811} taken from the list of
\cite{bedding+20}. As shown in Fig.\ref{HD}, the two stars show their
main excited modes in the range of 15-40 c/d and coincidentally both
show a strongly excited mode at $f\sim15.9$~c/d. Not much is known about
HD220811: It is a double star with a tiny separation of 0.4\arcsec. The
Transiting Exoplanet Survey Satellite (TESS) input catalogue says its
effective temperature is 7527K, which is close to the average Altair's
effective temperature of 7550K \citep{erspamer_north03}. Its $V\sin
i\sim$ 270~km/s, according to \cite{bedding+20}, is compatible with the
313~km/s equatorial velocity of Altair \citep{bouchaud+20}. The TESS
input catalogue also mentions a Hipparcos parallax of 8.57 mas and a
visual magnitude of 6.91. Using Altair's parallax 194.44 mas and visual
magnitude of 0.76, we deduce that the apparent luminosity of HD220811
is slightly larger than that of Altair, namely L$_{\rm HD220811}\simeq
1.8$L$_{\rm Altair}$. Since this luminosity is only the apparent one,
which depends on the inclination of the rotation axis, the two stars
may well be very similar.

The probable similarity of Altair and HD220811 as far as their oscillation spectrum and spectral type are concerned, together with their very fast rotation ($V_{\rm eq}\sim 300$~km/s), may illustrate the result of \cite{reese+16} that fast rotation reduces the number of excited modes in a star.
The parallel study of these two stars, and others with the same features, will be very useful to better appreciate all the effects of rotation.

\subsection{Rotation}

From interferometric and spectroscopic data, \cite{bouchaud+20} derived the angular velocity of Altair which corresponds to $\nu_{\rm rot}=3.08\pm0.03$~c/d. In fact, from the \cite{bouchaud+20} model of Altair, we have an idea of the surface differential rotation of this star. It is shown in Fig.~\ref{fig:Diff_rot} along with its uncertainty. A flux pattern at the surface of Altair may therefore show up at a frequency in between 2.9 and 3.16 c/d. A close-up of the spectrum in this range (see Fig.~\ref{fig:Rotation}) shows a signal at 2.99 c/d in the 2012 data set, which may be a signature of a such a flux pattern. Since Altair is known to be magnetically active \citep{robrade+09}, flux modulation by some magnetic feature at its surface is not impossible. However, this possible detection of rotation needs to be confirmed.

\subsection{Time variability}\label{tv}

The data sets collected by MOST over the years offer the possibility
to investigate the time variation of the modes
amplitude. However, the determination of the amplitude of a mode depends
on the determination of its frequency. The length of the light curves
and the time windows are unfortunately not constant from one set to
another. Mode frequencies are therefore more or less well determined
according to the data set at hand. However, for the modes listed in
Tab.~\ref{freq_altair}, frequencies can be considered as constant within
their error bar. Thus, we shall assume their constancy in time. We
therefore determined the best frequency of the main modes using all
data and projected the desired subset of the light curve on the simple
Fourier basis (e.g. Eq. \ref{the_model}) using least-squares to obtain
the time evolution of amplitudes over the years.


A first view of these variations of the main modes over the years is shown in Fig.~\ref{fig:Amp_vs_time}. The most prominent mode at f=15.7679 c/d shows a rather important growth (a factor of 1.5) between 1999 and 2007. With the origin of these variations being numerous (see \citealt{guzik+16} for a rather exhaustive list), a hint may be given by the timescales shown by these fluctuations. Fig.~\ref{fig:Amp_vs_time} shows that years should be considered, but phenomena such as convection may impose much shorter timescales and suggest for one to inspect variations over days. To this end, we concentrated on the 2012 data set, which is certainly the most appropriate with its 33 days length and its rather low noise. We first carried out a time-frequency analysis with a one-day time step and a 3.3 days interval. The result is shown in Fig.~\ref{fig:tfd}. 

We clearly see amplitude oscillations near 15.8 c/d due to the two beating modes at 15.7679 and 15.983 c/d, but an evolving trend is also visible along the time interval. This evolution is better seen if we split the time interval into disjoint windows of 3.3 days.
Shorter windows do not provide more information and they are influenced by noise or secondary window peaks too much. This is why we do not show the amplitude variation of the $f_7$-mode, which is too close to the main $f_6$-mode at f=15.7679~c/d. Hence, the amplitude variations of $f_6$ also include those of $f_7$.
Figure~\ref{fig:Amp12_vs_time} shows the amplitude variations of the five most prominent modes.
This figure suggests an amplitude variation on a timescale of 15 days for the main mode at $f_6$ and for the low amplitude mode at $f_{11}$, whose variations are anti-correlated with those of $f_6$. The amplitude of the beating effect between $f_6$ and $f_7$ is also shown in Fig.~\ref{fig:Amp12_vs_time} (black dashed line). For that, we generated a light curve where the two modes $f_6$ and $f_7$ have a constant amplitude, and we processed this artificial light curve in a similar way as the real one. We can clearly see that the beating effect produces amplitude variations that are much smaller than those observed on $f_6$.
Mode $f_{10}$ also seems to show, mildly, an amplitude evolution on  a 15-day timescale. The low-frequency mode at $f_5=3.527$~c/d seems to show an amplitude modulation on a 10-day timescale (e.g. Fig.~\ref{fig:tfd}), while the mode at $f_{12}$ also shows amplitude variations, but on a timescale similar to the window sampling, which hinders any conclusion being made.
The driving of $\delta$-Scuti oscillations is classically
attributed to the $\kappa$-mechanism associated with the opacity
bump generated by the partial second ionisation of helium around
50,000K \citep{baglin+73,balona+15}. However, such an opacity bump
also destabilises a whole layer of the star where thermal convection
arises. {This convection occurs at high Reynolds numbers and thus
excites a wide range of time and length scales. Using \cite{bouchaud+20}
Altair's model, we found that timescales of 10 or 15 days are well
within the possible timescales of convection} in the $\kappa$-exciting
layer. This coincidence may not be just by chance and further work is
needed to explain the way oscillations may be modulated on this timescale.

\section{Conclusions}

\cite{buzasietal05} discovered that Altair is the brightest $\delta$-Scuti
of the sky, and we
confirm this. The oscillation spectrum
of Altair now includes six new frequencies and we extend the range of
frequencies from 0.57 c/d to 35.82 c/d. The MOST observations of Altair,
distributed over several years, allowed us to bring the variations of the
mode amplitudes to light. The time frequency analysis of the 2012 data
set, which spans 33 days, showed characteristic timescales of $\sim$10
to 15 days, which can be imposed by the convective  layer associated
with the opacity bump of helium second ionisation. Since this opacity
bump is also known to drive oscillations in $\delta$-Scuti stars, a
good modelling of the coupling of thermal convection with oscillations,
in the spirit of \citet{GD11a}, would offer an opportunity to use these
amplitude variations to constrain the physics of the stars in this layer.

The number of modes excited at a detectable amplitude in Altair is still rather modest (15 if we were to gather all data). HD220811, which is quite similar to Altair as far as \teff\ and Vsin{\it i} are concerned, also shows a similar number of frequencies. This point is reminiscent of the theoretical (preliminary) result of \cite{reese+16}, which shows that as the rotation rate increases, mode excitation is less and less efficient. In other words, rotation tends to stabilise modes otherwise destabilised by the $\kappa$-mechanism.

Finally, we may have detected the signature of Altair's rotation as a
spectral peak popping up at 2.99 c/d, which is fully compatible with
present models of Altair \citep{bouchaud+20}. However, this detection
still demands confirmation as it is only visible in the 2012 light
curve.

The foregoing results encourage us to obtain more data on the seismology
of Altair. This will be the case with the shortly awaited analysis of
line profile variations that allow us to detect eigenmodes with a high
azimuthal wavenumber propagating in longitude (Rieutord et al. 2021 in
preparation). With a larger set of eigenfrequencies, we will be in a
better position to bring new constraints on the fundamental parameters
of Altair, our nearest fast rotating star.

\begin{acknowledgements}
We would like to warmly thank John Monnier for his kind reading of an
early version of the manuscript. CLD and MR acknowledge the support
of the French Agence Nationale de la Recherche (ANR), under grant ESRR
(ANR-16-CE31-0007-01), which made this work possible. SC acknowledges
financial support from the Centre National d’Études Spatiales (CNES,
France) and from the Agence Nationale de la Recherche (ANR, France)
under grant ANR-17-CE31-0018. We are also very grateful to all our
colleagues who made the MOST mission successful and left its precious
collection of data in the public domain.

\end{acknowledgements}


\begin{appendix}\label{AppendixA}
\section{Pre-processing of MOST data}

\begin{table}[t]
    \centering
    \begin{tabular}{ccccc}
    \hline
    &&&& \\
    &2007 & 2011 & 2012 & 2013 \\
From & 25 June & 26 July & 12 July & 17 July \\ 
{\small JD-JD2000} & 2732.18 & 4223.51 & 4576.09 & 4946.06\\
to   & 15 July & 30 July & 28 August & 25 July\\
{\small JD-JD2000} & 2752.34 & 4227.50 & 4609.49 & 4954.04\\
Duration (d) & 20.166 & 3.992 & 33.398 & 7.986 \\
Gaps &  /   & / &    1 d  & 3.5 d \\
N$_0$ & 37,729 & 4,834 & 23,555 & 4,849 \\
N  & 30,544 & 4,578 & 21,701 & 4,658 \\
Percentage of & \multirow{2}{*}{19.15\%} & \multirow{2}{*}{5.30\%} & \multirow{2}{*}{4.79\%} &  \multirow{2}{*}{3.94\%} \\
pointing errors &&&&\\
\hline
    \end{tabular}
    \bigskip
    \caption{Summary of characteristics of data sets. We note that N$_0$ is the number of pictures in the initial series, while $N$ is the number of pictures left after the removal of the problematic ones.}
    \label{tab:the_data}
\end{table}

\begin{figure}[t]
    \centering
    \includegraphics[width=1.0\linewidth]{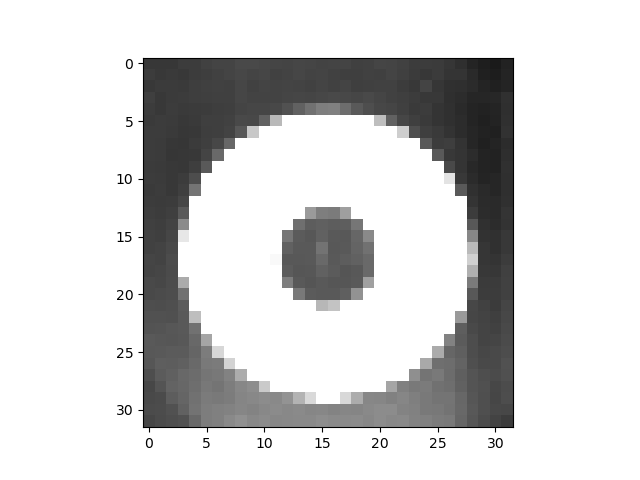}
    \caption{Example of a Fabry image obtained from observations of Altair with the Fabry lens from the 2012 data set. We note the non-uniform background due to stray light.}
    \label{fig:fabry-image}
\end{figure}

Data from MOST come as a series of Fabry images where the star light is
projected onto the CCD by a Fabry microlens as an annulus covering about
450 pixels as shown in Fig \ref{fig:fabry-image}.  We reduced the data
using different techniques inspired by  \cite{reegen+06}.

\begin{figure}[ht]
    \centering
    \includegraphics[width=1.0\linewidth]{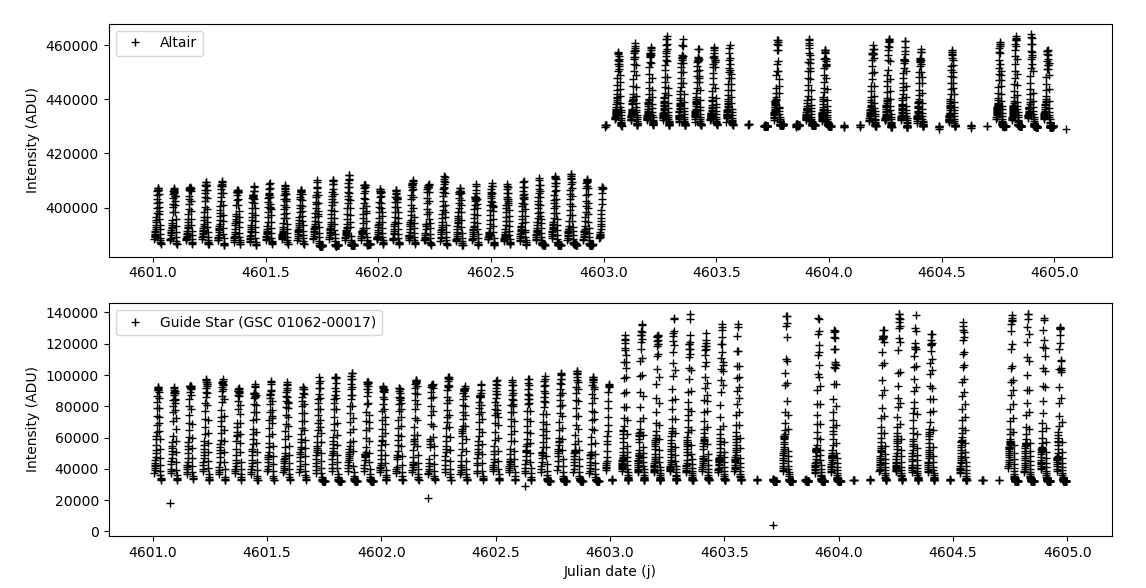}
    \caption{Mean intensity versus time (2012 data set) for Altair (top) and the guide star (bottom).}
    \label{fig:discontinuity}
\end{figure}

The first step was to  remove discontinuities that appear in the time series of Altair.
 These discontinuities are also present in the time series of the guide stars (fig. \ref{fig:discontinuity}). By analysing the headers in the fits file, we believe that these discontinuities are due to a change in the acquisition parameters of the telescope to improve the contrast. We removed these discontinuities by adding the appropriate constant.

The Attitude Control System gives a set of xy-errors  for each image. These error values indicate whether or not the target is inside the nominal Fabry lens area.
When the x-errors or y-errors cross the limits of the interval $[-25.8, 25.8]$ (in arcsec) during the integration time, the maximum value is returned. We rejected all
images with out-of-range xy-errors and $3\sigma$ outliers.

The effects of removing bad data are shown in Tab.~\ref{tab:the_data}. For the 2007 data, this sorting removes \mbox{$\sim19\%$} of the data points. This rather high rate of bad data was related to the pointing precision of MOST, which was improved later, and does not affect data from 2011, 2012, and 2013.

The last step needed to derive the light curves of Altair was to decorrelate Altair's light from stray light. This important step is described in the main text.

\end{appendix}

\bibliographystyle{aa}
\bibliography{bibnew}

\begin{thebibliography}{29}
\expandafter\ifx\csname natexlab\endcsname\relax\def\natexlab#1{#1}\fi

\bibitem[{{Baglin} {et~al.}(1973){Baglin}, {Breger}, {Chevalier}, {Hauck}, {Le
  Contel}, {Sareyan}, \& {Valtier}}]{baglin+73}
{Baglin}, A., {Breger}, M., {Chevalier}, C., {et~al.} 1973, A\&A, 23, 221

\bibitem[{{Balona}(2014)}]{balona14}
{Balona}, L.~A. 2014, MNRAS, 439, 3453

\bibitem[{{Balona} {et~al.}(2015){Balona}, {Daszy{\'n}ska-Daszkiewicz}, \&
  {Pamyatnykh}}]{balona+15}
{Balona}, L.~A., {Daszy{\'n}ska-Daszkiewicz}, J., \& {Pamyatnykh}, A.~A. 2015,
  MNRAS, 452, 3073

\bibitem[{{Bedding} {et~al.}(2020){Bedding}, {Murphy}, {Hey}, {Huber}, {Li},
  {Smalley}, {Stello}, {White}, {Ball}, {Chaplin}, {Colman}, {Fuller},
  {Gaidos}, {Harbeck}, {Hermes}, {Holdsworth}, {Li}, {Li}, {Mann}, {Reese},
  {Sekaran}, {Yu}, {Antoci}, {Bergmann}, {Brown}, {Howard}, {Ireland},
  {Isaacson}, {Jenkins}, {Kjeldsen}, {McCully}, {Rabus}, {Rains}, {Ricker},
  {Tinney}, \& {Vanderspek}}]{bedding+20}
{Bedding}, T.~R., {Murphy}, S.~J., {Hey}, D.~R., {et~al.} 2020, Nature, 581,
  147

\bibitem[{{Bouchaud} {et~al.}(2020){Bouchaud}, {Domiciano de Souza},
  {Rieutord}, {Reese}, \& {Kervella}}]{bouchaud+20}
{Bouchaud}, K., {Domiciano de Souza}, A., {Rieutord}, M., {Reese}, D.~R., \&
  {Kervella}, P. 2020, A\&A, 633, A78

\bibitem[{{Buzasi} {et~al.}(2005){Buzasi}, {Bruntt}, {Bedding}, {Retter},
  {Kjeldsen}, {Preston}, {Mandeville}, {Suarez}, {Catanzarite}, {Conrow}, \&
  {Laher}}]{buzasietal05}
{Buzasi}, D.~L., {Bruntt}, H., {Bedding}, T.~R., {et~al.} 2005, ApJ, 619, 1072

\bibitem[{{Charpinet} {et~al.}(2010){Charpinet}, {Green}, {Baglin}, {Van
  Grootel}, {Fontaine}, {Vauclair}, {Chaintreuil}, {Weiss}, {Michel},
  {Auvergne}, {Catala}, {Samadi}, \& {Baudin}}]{charpinet2010}
{Charpinet}, S., {Green}, E.~M., {Baglin}, A., {et~al.} 2010, \aap, 516, L6

\bibitem[{{Chen} {et~al.}(2016){Chen}, {Li}, {Lai}, \& {Wu}}]{chen+16}
{Chen}, X.~H., {Li}, Y., {Lai}, X.~J., \& {Wu}, T. 2016, A\&A, 593, A69

\bibitem[{{Domiciano de Souza} {et~al.}(2005){Domiciano de Souza}, Kervella,
  Jankov, Vakili, Ohishi, Nordgren, \& Abe}]{domiciano+05}
{Domiciano de Souza}, A., Kervella, P., Jankov, S., {et~al.} 2005, A\&A, 442,
  567

\bibitem[{{Erspamer} \& {North}(2003)}]{erspamer_north03}
{Erspamer}, D. \& {North}, P. 2003, A\&A, 398, 1121

\bibitem[{{Espinosa Lara} \& {Rieutord}(2013)}]{ELR13}
{Espinosa Lara}, F. \& {Rieutord}, M. 2013, A\&A, 552, A35

\bibitem[{{Garc{\'{\i}}a Hern{\'a}ndez} {et~al.}(2015){Garc{\'{\i}}a
  Hern{\'a}ndez}, {Mart{\'{\i}}n-Ruiz}, {Monteiro}, {Su{\'a}rez}, {Reese},
  {Pascual-Granado}, \& {Garrido}}]{garciahernandez_etal15}
{Garc{\'{\i}}a Hern{\'a}ndez}, A., {Mart{\'{\i}}n-Ruiz}, S., {Monteiro},
  M.~J.~P.~F.~G., {et~al.} 2015, ApJ Lett., 811, L29

\bibitem[{{Gastine} \& {Dintrans}(2011)}]{GD11a}
{Gastine}, T. \& {Dintrans}, B. 2011, A\&A, 528, A6

\bibitem[{{Guzik} {et~al.}(2016){Guzik}, {Kosak}, {Bradley}, \&
  {Jackiewicz}}]{guzik+16}
{Guzik}, J.~A., {Kosak}, K., {Bradley}, P.~A., \& {Jackiewicz}, J. 2016, IAU
  Focus Meeting, 29B, 560

\bibitem[{{Huber} \& {Reegen}(2008)}]{huber_reegen08}
{Huber}, D. \& {Reegen}, P. 2008, Comm. in Asteroseismology, 152, 77

\bibitem[{{Lenz} \& {Breger}(2005)}]{lenz_breger05}
{Lenz}, P. \& {Breger}, M. 2005, Comm. in Asteroseismology, 146, 53

\bibitem[{{Monnier} {et~al.}(2010){Monnier}, {Townsend}, {Che}, {Zhao},
  {Kallinger}, {Matthews}, \& {Moffat}}]{monnier_etal10}
{Monnier}, J.~D., {Townsend}, R.~H.~D., {Che}, X., {et~al.} 2010, ApJ, 725,
  1192

\bibitem[{{Monnier} {et~al.}(2007){Monnier}, {Zhao}, {Pedretti}, {Thureau},
  {Ireland}, {Muirhead}, {Berger}, {Millan-Gabet}, {Van Belle}, {ten
  Brummelaar}, {McAlister}, {Ridgway}, {Turner}, {Sturmann}, {Sturmann}, \&
  {Berger}}]{monnier_etal07}
{Monnier}, J.~D., {Zhao}, M., {Pedretti}, E., {et~al.} 2007, Science, 317, 342

\bibitem[{{Montgomery} \& {O'Donoghue}(1999)}]{montgomery+99}
{Montgomery}, M.~H. \& {O'Donoghue}, D. 1999, Delta Scuti Star Newsletter, 13,
  28

\bibitem[{{Poretti} {et~al.}(2009){Poretti}, {Michel}, {Garrido},
  {Lef{\`e}vre}, {Mantegazza}, {Rainer}, {Rodr{\'\i}guez}, {Uytterhoeven},
  {Amado}, {Mart{\'\i}n-Ruiz}, {Moya}, {Niemczura}, {Su{\'a}rez}, {Zima},
  {Baglin}, {Auvergne}, {Baudin}, {Catala}, {Samadi}, {Alvarez}, {Mathias},
  {Papar{\`o}}, {P{\'a}pics}, \& {Plachy}}]{poretti+09}
{Poretti}, E., {Michel}, E., {Garrido}, R., {et~al.} 2009, A\&A, 506, 85

\bibitem[{{Reegen} {et~al.}(2006){Reegen}, {Kallinger}, {Frast}, {Gruberbauer},
  {Huber}, {Matthews}, {Punz}, {Schraml}, {Weiss}, {Kuschnig}, {Moffat},
  {Walker}, {Guenther}, {Rucinski}, \& {Sasselov}}]{reegen+06}
{Reegen}, P., {Kallinger}, T., {Frast}, D., {et~al.} 2006, \mnras, 367, 1417

\bibitem[{{Reese} {et~al.}(2017){Reese}, {Dupret}, \& {Rieutord}}]{reese+16}
{Reese}, D.~R., {Dupret}, M.-A., \& {Rieutord}, M. 2017, in European Physical
  Journal Web of Conferences, Vol. 160, European Physical Journal Web of
  Conferences, 02007

\bibitem[{{Reese} {et~al.}(2012){Reese}, {Marques}, {Goupil}, {Thompson}, \&
  {Deheuvels}}]{reese+12}
{Reese}, D.~R., {Marques}, J.~P., {Goupil}, M.~J., {Thompson}, M.~J., \&
  {Deheuvels}, S. 2012, A\&A, 539, A63

\bibitem[{{Rieutord} {et~al.}(2016){Rieutord}, {Espinosa Lara}, \&
  {Putigny}}]{RELP16}
{Rieutord}, M., {Espinosa Lara}, F., \& {Putigny}, B. 2016, J. Comp. Phys.,
  318, 277

\bibitem[{{Robrade} \& {Schmitt}(2009)}]{robrade+09}
{Robrade}, J. \& {Schmitt}, J.~H.~M.~M. 2009, A\&A, 497, 511

\bibitem[{{van Belle} {et~al.}(2001){van Belle}, {Ciardi}, {Thompson},
  {Akeson}, \& {Lada}}]{vanbelle+01}
{van Belle}, G.~T., {Ciardi}, D.~R., {Thompson}, R.~R., {Akeson}, R.~L., \&
  {Lada}, E.~A. 2001, ApJ, 559, 1155

\bibitem[{{van Leeuwen}(2007)}]{vanleeuwen07}
{van Leeuwen}, F. 2007, A\&A, 474, 653

\bibitem[{{Walker} {et~al.}(2003){Walker}, {Matthews}, {Kuschnig}, {Johnson},
  {Rucinski}, {Pazder}, {Burley}, {Walker}, {Skaret}, {Zee}, {Grocott},
  {Carroll}, {Sinclair}, {Sturgeon}, \& {Harron}}]{walker+03}
{Walker}, G., {Matthews}, J., {Kuschnig}, R., {et~al.} 2003, Pub. Astron. Soc.
  Pacific, 115, 1023

\bibitem[{{Zong} {et~al.}(2016){Zong}, {Charpinet}, \& {Vauclair}}]{zong2016}
{Zong}, W., {Charpinet}, S., \& {Vauclair}, G. 2016, \aap, 594, A46

\end{thebibliography}

\end{document}